\documentclass[prx,10pt,twocolumn,superscriptaddress,floatfix,showpacs]{revtex4-1}

\usepackage[british]{babel}  
\usepackage[scaled=0.86]{berasans}  
\usepackage[colorlinks=true, allcolors=blue, urlcolor=blue]{hyperref}  
\usepackage{graphicx} 
\usepackage[babel]{microtype}  
\usepackage{amsmath,amssymb,amsthm,bm,amsfonts,mathrsfs,bbm} 

\usepackage{xspace}  
\usepackage{pgfplots}
\usepackage{xcolor,colortbl}
\usepackage{array}
\usepackage{bigstrut}
\usepackage{pifont}
\newcommand{\cmark}{\ding{51}}%
\newcommand{\xmark}{\ding{55}}%

\newcommand{\tr}{\text{tr}}

\newcommand{\be}{\begin{equation}}
\newcommand{\ee}{\end{equation}}
\newcommand{\bea}{\begin{eqnarray}}
\newcommand{\eea}{\end{eqnarray}}
\newcommand{\bes}{\begin{equation*}}
\newcommand{\ees}{\end{equation*}}
\newcommand{\beas}{\begin{eqnarray*}}
\newcommand{\eeas}{\end{eqnarray*}}

\def\<{\langle}
\def\>{\rangle}

\def\tr{\mathrm{tr}}

\newtheorem{Proposition}{Proposition}

\theoremstyle{definition}
\newtheorem{ex}{Example}

\newtheorem{defn}{Definition}
\newtheorem{remark}{Remark}

\newtheorem*{lipschitzLem*}{Lemma \ref{lipschitz}}
\newtheorem*{lipschitzCubeLem*}{Lemma \ref{lipschitzCube}}
\newtheorem*{pgmNearlyOptimalThm*}{Theorem \ref{pgmNearlyOptimal}}

\usepackage{amsfonts}
\usepackage{amssymb}
\usepackage{mathrsfs}
\usepackage{amsthm}
\usepackage{graphicx}
\usepackage{amsmath}
\usepackage{color}
\usepackage{tikz}
\newcommand{\cc}{\mathbb{C}}

\newcommand{\bei}{\begin{itemize}}
\newcommand{\eei}{\end{itemize}}

\def\oper{{\mathchoice{\rm 1\mskip-4mu l}{\rm 1\mskip-4mu l}{\rm 1\mskip-4.5mu l}{\rm 1\mskip-5mu l}}}

\begin{document}

\title{On the structure of mirrored operators obtained from optimal entanglement witnesses}

\author{Anindita Bera}
\affiliation{Institute of Physics, Faculty of Physics, Astronomy and Informatics,
Nicolaus Copernicus University, Grudzi\c{a}dzka 5/7, 87--100 Toru{\'n}, Poland}

\author{Joonwoo Bae}
\affiliation{School of Electrical Engineering, Korea Advanced Institute of Science and Technology (KAIST),
291 Daehak-ro, Yuseong-gu, Daejeon 34141, Republic of Korea}

\author{Beatrix C. Hiesmayr}
\affiliation{University of Vienna, Faculty of Physics, W\"ahringerstrasse 17, 1090 Vienna, Austria}

\author{Dariusz Chru{\'s}ci{\'n}ski}
\affiliation{Institute of Physics, Faculty of Physics, Astronomy and Informatics,
Nicolaus Copernicus University, Grudzi\c{a}dzka 5/7, 87--100 Toru{\'n}, Poland}

\begin{abstract}
Entanglement witnesses (EWs) are a versatile tool in the verification of entangled states. The framework of mirrored EW  doubles the power of a given EW by introducing its twin -- a mirrored EW -- whereby two EWs related by mirroring can bound the set of separable states more efficiently. In this work, we investigate the relation between the EWs and its mirrored ones, and present a conjecture which claims that  the mirrored operator obtained from an optimal EW is either a positive operator or a decomposable EW, which implies that positive-partial-transpose entangled states, also known as the bound entangled states, cannot be detected. This conjecture is reached by studying numerous known examples of optimal EWs. However, the mirrored EWs obtained from the non-optimal ones can be non-decomposable as well. We also show that mirrored operators obtained from the extremal decomposable witnesses are positive semi-definite. Interestingly, the witnesses that violate the well known conjecture of Structural Physical Approximation, do satisfy our conjecture. The intricate relation between these two conjectures is discussed  and it reveals a novel structure of the separability problem.
\end{abstract}

\maketitle

\section{Introduction}

Entanglement witnesses (EWs) are a both theoretical and experimental tool to detect entangled states \cite{HHHH,EW1,EW2,TOPICAL,KYE,ani18}. When an entangled state $\rho$ realized in experiment is identified by quantum state tomography, there exists an EW that finds if it is entangled, i.e.,
\bea
\tr[W\rho]<0,~\mathrm{whereas}~ 0\leq \tr[W\sigma_{\mathrm{sep}}],~~\forall~\sigma_{\mathrm{sep}}\in\mathrm{SEP} ~~~ \label{eq:ew} 
\eea
where $\mathrm{SEP}$ denotes the set of separable states. In fact, Eq. (\ref{eq:ew}) can be used as a definition of entangled states: a bipartite state $\rho$ is entangled if and only if there exists an EW $W$ such that $\tr [W \rho] <0$, and therefore $W$ detects, i.e. \textit{witnesses} the entanglement  \cite{HHHH}. 


Since EWs correspond to the Hermitian operators, they can be realized experimentally for the verification of entangled states. This also means that entanglement can be directly verified in experiment without the identification of a given state, i.e. by in general, less measurement setups. In general, an EW can be decomposed into local observables, 
\bea
W = \sum_{i,j} c_{ij} A_i\otimes B_j, 
\label{eq:lo}
\eea
with some numbers of $c_{ij}$ and local observables $A_i$ and $B_j$. A collection of expectation values of local observables $\langle A_i \otimes B_j \rangle_{\rho} =  \tr[A_i\otimes B_j \rho]$ so that one computes $\tr [W \rho] = \sum_{ij} c_{ij} \langle A_i\otimes B_j\rangle_{\rho}$ { detects} if a state $\rho$ is entangled. 

Due to the well known Choi-Jamio{\l}kowski isomorphism \cite{Jam,MDChoi}, there is  one to one correspondence between the block-positive operators in $\mathcal{H}_A \otimes \mathcal{H}_B$ and positive maps $\mathcal{B}(\mathcal{H}_A) \to \mathcal{B}(\mathcal{H}_A)$, where $\mathcal{B}(\mathcal{H})$ denotes bounded linear operators acting on $\mathcal{H}$ (in this paper we consider only finite dimensional Hilbert spaces). Entanglement witnesses correspond to positive but not completely positive maps~\cite{Paulsen,Stormer,ani22}.


Optimal EWs are of particular importance \cite{Lew}. An EW $W$ is called optimal if $W - \epsilon P$ for all non-negative operators $P\geq 0$ and for any $\epsilon >0 $ is no longer an EW. That is, one cannot improve $W$ by subtracting a positive operator. 
Entanglement witnesses, being Hermitian operators, represent physical observables and hence in principle, can be implemented in the laboratory. However, positive maps which are not completely positive are not physically realizable. The idea of  structural physical approximation (SPA) is to 
mix a positive map with {an amount  of the completely depolarizing map as small as possible} in order to obtain a physically realizable completely positive map \cite{SPA-01,SPA-02}. Equivalently, SPA to an entanglement witness $W$ is defined by 
\begin{equation}  \label{SPA}
    X = p\; \oper_A \otimes \oper_B + W ,
\end{equation}
with the smallest $p>0$ such that $X \geq 0$ (i.e. $p = - \lambda_{\rm min}$, where $\lambda_{\rm min}$ is a minimal eigenvalue of $W$)

\begin{equation}  \label{SPA}
    X = p~ \oper_A \otimes \oper_B + W ,
\end{equation}
with the smallest $0<p<1$ such that $X \geq 0$. Note that  {a SPA operator} $X$ may also be interpreted as a not normalized quantum state. 

The SPA conjecture in Ref.~\cite{SPA-1} { has} asserted that SPA to an optimal EW leads to a separable state $X$, or, equivalently, SPA to optimal positive trace-preserving map leads to entanglement-breaking quantum channels (cf. also Refs.~\cite{SPA-2,SPA-3} and Ref.~\cite{SPA-4} for an review). It was firstly supported by many examples of optimal EWs~\cite{EX1,EX2,S71,SPA-4}, however, later it was disproved~\cite{hakyend,SPA-Stormer,dar-gni}.

In this paper, we consider a similar concept based on the notion of a mirrored operator introduced in Ref.~\cite{MEW}. Given an EW $W$, we define a mirrored operator by
\begin{equation}   \label{M!}
    W_{\rm M} = \mu \oper_A \otimes \oper_B - W, 
\end{equation}
with the smallest $\mu>0$ such that $W_{\rm M}$ is block-positive, i.e. $\< \psi\otimes \phi|W_{\rm M}|\psi \otimes \phi\> \geq 0$. Moreover, if the maximal eigenvalue of $W$ satisfies $\lambda_{\rm max} > \mu$, then $W_{\rm M}$ is an EW and hence one has a pair $(W,W_{\rm M})$ of mirrored EWs~\cite{MEW},  which can double up the capability of detecting entangled states. This framework is also referred to as ``\textit{entanglement witnesses 2.0}'' since every witness comes with another one, in analogy to software programs that improve with each new version.

An important property of EWs is its (non)--decomposibility. An EW is called decomposable if $W=A + B^\Gamma$, with $A,B \geq 0$ and $\Gamma$ stands for the partial transposition. 
Note that decomposable witnesses cannot detect PPT entangled states, i.e. if $W = A + B^\Gamma$, then ${\rm Tr}(W\sigma)\geq 0 $ for all PPT states $\sigma$. This is easily proved by exploiting the fact that the trace is invariant under transposition, i.e. for a PPT state $\sigma$, ${\rm Tr}(A\sigma)+{\rm Tr}( B^\Gamma\sigma)={\rm Tr}(A\sigma)+{\rm Tr}( B\sigma^\Gamma) \geq 0$, since $A,B\geq 0$ and $\sigma,\sigma^\Gamma \geq 0$.

In this paper, we investigate the structure of mirrored EWs. In particular, we address the following question: given an optimal EW, what are the properties of the corresponding mirrored one? As an answer, we propose a conjecture which says that given an optimal EW, its mirror operator $W_{\rm M}$ is either a decomposable EW or just a positive operator. In other words, there does not exist a mirrored pair of non-decomposable EWs $(W,W_{\rm M})$ such that at least one of them is optimal. The assumption about optimality is crucial. Actually, we show one can construct  a mirrored pair of non-decomposable EWs but none of them is optimal. We believe that our analysis paves a new avenue of finding a fine structure of the set of entanglement witnesses { and thus the structure of separable and PPT-entangled states in the Hilbert space}.

The paper is organized as follows. In Sec.~\ref{sec:MEW}, we discuss the concept of mirrored EWs and propose our conjecture. In Sec.~\ref{SEC-EW}, we review the basic properties of optimal entanglement witnesses. Sec.~\ref{SEC-D} provides the analysis of our conjecture for decomposable EWs. 
Moreover, we show that our conjecture holds true for a class of extremal decomposable EWs. 
In Section \ref{SEC-ND}, we provide several examples of optimal non-decomposable EWs supporting the above conjecture. Additionally, we construct an entanglement witness in $\mathbb{C}^4 \otimes \mathbb{C}^4$ which is non-decomposable and not optimal and find that the mirrored EW is non-decomposable as well. Interestingly, we include the analysis of optimal EWs which were used to disprove SPA conjecture \cite{hakyend,dar-gni}. Numerical analysis shows that both {examples} support our conjecture. We finally conclude in Sec.~\ref{SEC-C}.

\section{  Mirrored Entanglement witnesses }
\label{sec:MEW}

\begin{figure}
    \centering
    \includegraphics[width=8.8cm]{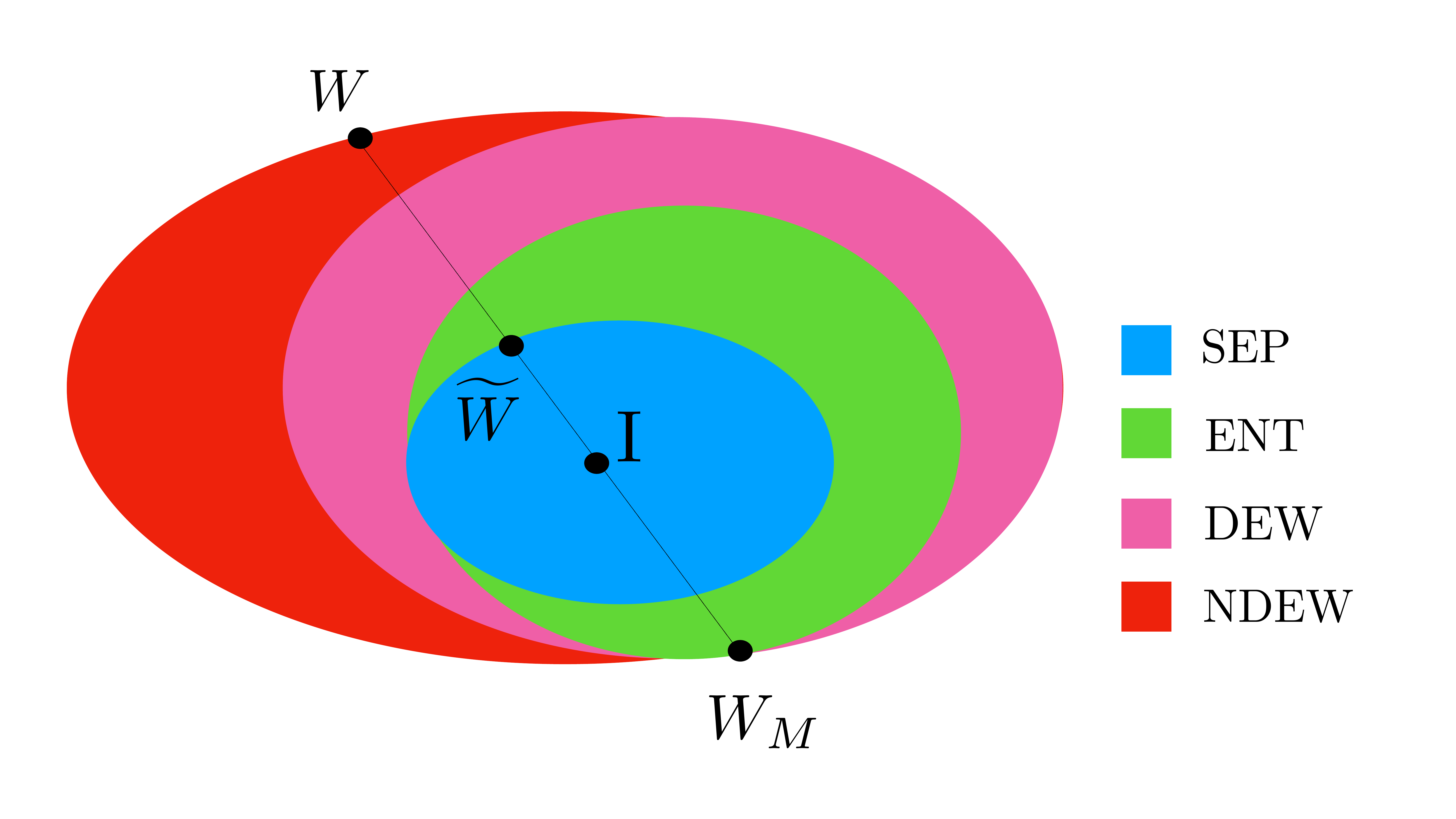}
    \caption{The sets of separable states (SEP), entangled states (ENT), decomposable EWs (DEWs), and non-decomposable EWs (NDEWs) are compared in the view of our conjecture and the SPA one. Given an NDEW $W$, SPA to $W$ is denoted by $\widetilde{W}$ and $W_M$ is the corresponding mirrored witness.
    The SPA conjecture addresses that SPA to { an optimal} EWs are separable states. Counterexamples are, however, obtained in Refs.~\cite{hakyend, SPA-Stormer, dar-gni}. Our conjecture suggests that a mirrored operator $W_\mathrm{M}$ obtained from an optimal EW is either a decomposable EW or a positive operator. 
    }
    \label{conjecture}
\end{figure}

The concept of mirrored EW is closely related to SPA  which we shortly summarize here (see Fig.~\ref{conjecture}). Given a   bipartite operator $Q\geq 0$, let us define the followings:
\begin{eqnarray}
    a_- &:=& \inf_{|\psi \otimes \phi\rangle} \,  \langle \psi \otimes \phi| { Q }| \psi \otimes \phi \rangle, \nonumber \\
    a_+ &:=& \sup_{|\psi \otimes \phi\rangle} \,  \langle \psi \otimes \phi| { Q }| \psi \otimes \phi \rangle,
\end{eqnarray}
where $|\psi \otimes \phi \rangle$ is normalized product vector in $\mathcal{H}_A \otimes \mathcal{H}_B$. 
Defining a pair of block-positive operators with the values obtained above
\begin{eqnarray}
  \label{WW}
    W_- &:=& a_+ \oper_A \otimes \oper_B - Q  , \nonumber \\ 
    W_+ &:=& Q  - a_- \oper_A \otimes \oper_B ,
\end{eqnarray}
or equivalently
\begin{equation}
W_+ + a_- \oper_A \otimes \oper_B = Q = a_+ \oper_A \otimes \oper_B - W_- \ , 
\end{equation}
one can say that a positive operator $Q$ represents two complementary SPA $W_+$ and $W_-$, i.e a {\em positive } SPA $W_+$ and {\em negative} SPA to $W_-$.

Moreover, the paired EWs in \eqref{WW} are related as follows:
\begin{equation}
W_+ +  W_-=  \mu \oper_A \otimes \oper_B, 
\label{eq:rel}
\end{equation}
where $\mu = a_+ - a_-$.
Hence, we can say that two EWs $W_+$ and $W_-$ are mirrored { to} each other, see also Eqs. (\ref{SPA}) and (\ref{M!}). To clarify the above relations, the separability bounds of $W_+$ {are}
\bea
0\leq \tr[W_+ \sigma_{\mathrm{sep}}] \leq \mu, ~~\forall \sigma_{\mathrm{sep}}\in \mathrm{SEP}, \label{eq:w+}
\eea
where the upper bound is equivalent to the condition $\tr[ W_- \sigma_{\mathrm{sep}} ]\geq 0$, { the very definition of an entanglement witness}. From Eq. (\ref{eq:rel}), a reciprocal relation leads to the separability bounds for $W_-$,
\bea
0\leq \tr[W_- \sigma_{\mathrm{sep}}] \leq \mu, ~~\forall \sigma_{\mathrm{sep}}\in \mathrm{SEP}, \label{eq:w-}
\eea
where the upper bound is equivalent to the condition $\tr[ W_+ \sigma_{\mathrm{sep}} ]\geq 0$. 

Having introduced mirrored EWs, we are now ready to address the conjecture.\\
\\



    {\bf Conjecture:}~~
{\em {A} mirrored operator { obtained from} an optimal EW is either a positive operator or a decomposable EW { and hence cannot detect PPT-entangled states}.} \\
\\

Equivalently, the above conjecture asserts that there does not exist a pair of non-decomposable EWs $(W,W_{\rm M})$ such that at least one of them is optimal. Similarly to the SPA conjecture, the conjecture mentioned above is concerned with the optimality of an EW and its mirrored one.  The conjecture is motivated by  the observation in Eq.~(\ref{eq:rel}) that shows a trade-off relation between EWs: if one is closer to the identity, the other  one is further away from the identity.  Our conjecture is that this observation is related to the optimality of EWs. 



\section{Optimal Entanglement witnesses: Basic properties }  \label{SEC-EW}

Given an EW $W$, let us denote $\mathcal{D}_{W}$, by a subset of states detected by $W$ \cite{Lew}, i.e. a set of states $\rho$ such that  $\tr(W\rho) < 0$. One calls $W_1$ is finer than $W_{2}$ if $ D_{W_{2}} \subseteq D_{W_{1}}$. 
 $W$ is optimal if there is no finer EW than $W$. Optimality of $W$ is equivalent to the following property  \cite{Lew}: if $W$ is optimal, then $W-P$ is no longer an EW, where $P$ is an arbitrary positive operator.  
It means that one cannot improve $W$ (i.e. make it finer)  by subtracting $P \geq 0$. Note that optimality does not protect to subtract a block-positive operator. Finally, $W$ is extremal if and only if $W-B$ is no longer an EW, where $B$ is an arbitrary block- positive operator such that $B \neq \lambda W$.   

Clearly, any extremal EW is optimal. However, the converse needs not be true. An EW corresponding to so-called reduction map $R_n : M_n(\mathbb{C}) \to M_n(\mathbb{C}) $ defined by
\begin{equation} \label{Rn}
    R_n(X) = \oper_n \tr X - X,  
\end{equation}
is optimal for all $n\geq 2$ but extremal only for $n=2$. 

In general, given $W$ it is very hard to check whether it is optimal. There exists, however, an operational  sufficient condition for optimality \cite{Lew}. Denote by $P_W$ a set of product vectors $|\psi \otimes \phi\rangle$ such that
\begin{equation}
    \< \psi \otimes \phi |W| \psi \otimes \phi \> = 0 .
\end{equation}
One has the following \cite{Lew}
\begin{Proposition} If ${\rm span}\, P_W = \mathcal{H}_A \otimes \mathcal{H}_B$, then $W$ is optimal. 
\end{Proposition}
In this case, i.e. when ${\rm span}~P_W = \mathcal{H}_A \otimes \mathcal{H}_B$, one says that $W$ has the spanning property. It should be stressed, however, that there exists optimal EWs without spanning property (cf. recent discussion in Ref.~\cite{LAA}). 

Consider a decomposable EW $W = A + B^\Gamma$ in $\mathbb{C}^n \otimes \mathbb{C}^m$. Recall that $W$ is optimal if $W = B^\Gamma$ and $B$ is supported on completely entangled subspace (CES) \cite{maciej00}. A linear subspace $\Sigma \subset \mathbb{C}^n \otimes \mathbb{C}^m$ defines a CES if it does not contain a product vector. It is well known that a maximal dimension of any CES in  $\mathbb{C}^n \otimes \mathbb{C}^m$ is $(n-1)(m-1)$ \cite{CES1,CES2} The simplest example of CES is a 1-dimensional subspace spanned by an arbitrary entangled vector $|\Psi\rangle \in \mathbb{C}^n \otimes \mathbb{C}^m$. The corresponding entanglement witness $|\Psi\>\<\Psi|^\Gamma$ is extremal \cite{TOPICAL}. It is, therefore, clear that any decomposable EW is a convex combination of extremal witnesses. 

For non-decomposable EWs, the situation is much more complicated \cite{jpa22}. Recall that a bipartite state is called a PPT state (Positive Partial Transpose) if $\rho^\Gamma \geq 0$, i.e. both $\rho$ and $\rho^\Gamma$ are legitimate quantum states. Now, $W$ is a non-decomposable EW if and only if it detects a PPT-entangled state. Let $\mathcal{D}^{\rm PPT}_W$ be a set of PPT states detected by $W$. 
Now,  $W_1$ is non-decomposable--finer than $W_2$ if  $\mathcal{D}^{\rm PPT}_{W_2} \subseteq \mathcal{D}^{\rm PPT}_{W_1}$.  
An EW $W$ is non-decomposable--optimal if there is no non-decomposable--finer EW than $W$. Actually, if an EW $W$ is non-decomposable--optimal, then $W-D$ for a PPT operator $D$ is no longer an EW. 
It means that one cannot improve $W$ (i.e. make it finer)  by subtracting a PPT operator $P$. Interestingly, it has been proven~\cite{Lew}

\begin{Proposition}
$W$ is non-decomposable--optimal if and only if both $W$ and $W^\Gamma$ are optimal.  
\end{Proposition}
A similar concept but on the level of states is provided by so-called {\em edge state} \cite{Lew-2}. 

\begin{figure}
    \centering
    \includegraphics[width=8.3cm]{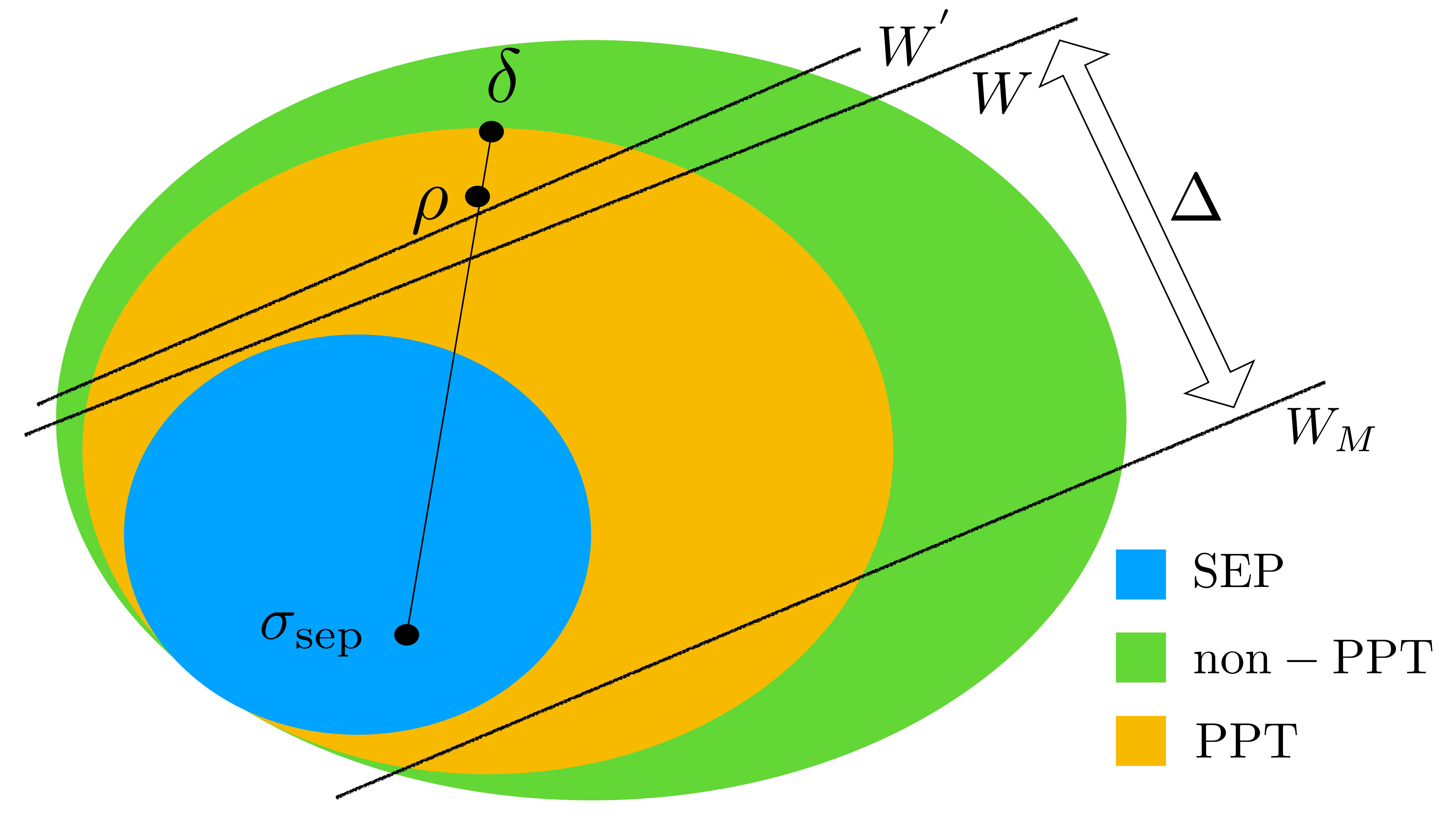}
    \caption{ The set of quantum states is depicted. A PPT-entangled state $\rho$ can be expressed as a convex mixture of a separable state $\sigma_{\mathrm{sep}}$ and an edge state $\delta$. EWs such as $W$ or $W^{'}$, which are also non-decomposable, may detect an entangled state $\rho$. For mirrored EWs $W$ and $W_{\mathrm{M}}$, the separability window is denoted by $\Delta$. }
    \label{qs}
\end{figure}

\begin{defn} A PPT-entangled state $\rho$ is called an edge state if $\sigma=\rho - \epsilon |\psi \otimes \phi\>\<\psi \otimes \phi|$ is no longer a PPT operator for arbitrary product state $|\psi \otimes \phi\rangle$ and arbitrarily small $\epsilon$, i.e. either $\sigma$ or $\sigma^\Gamma$ is not positive. 
\end{defn}
It simply means that if $\rho$ is an edge state, then one cannot subtract any PPT state out of it without destroying a PPT property, as it is shown in Fig. \ref{qs}.  Authors of \cite{Lew-2} provided the following representation of non-decomposable EWs: let $\rho_{\rm edge}$ be an edge state. To construct an EW detecting $\rho_{\rm edge}$, consider two positive operators $P$ and $Q$ such that
$$  {\rm Ran}\,P \subseteq {\rm Ker}\, \rho_{\rm edge}\ , \ \ \ \ {\rm Ran}\,Q \subseteq {\rm Ker}\, \rho_{\rm edge}^\Gamma, $$
where ${\rm Ran}$ and ${\rm Ker}$ denote the range and the kernel of the corresponding operator, respectively. Define
\begin{equation}  \label{WPQ}
W = P + Q^\Gamma - \epsilon_-\, \oper_A \otimes \oper_B ,    
\end{equation}
with $\epsilon_- =  \inf_{|\psi \otimes \phi\rangle} \,  \<\psi \otimes \phi| (P+Q^\Gamma)| \psi \otimes \phi \> $. By construction, $W$ is block-positive and
\begin{equation}
    \tr(W \rho_{\rm edge}) = \tr[(P+Q^\Gamma)\rho_{\rm edge}] - \epsilon_- = - \epsilon_- < 0 ,
\end{equation}
which shows that $W$ is a non-decomposable EW detecting $\rho_{\rm edge}$. Note that  using Eq.~(\ref{WPQ}), one can easily find the mirrored operator
\begin{equation}  \label{WPQM}
W_{\rm M} =  \epsilon_+\, \oper_A \otimes \oper_B  -    (P + Q^\Gamma) ,
\end{equation}
with $\epsilon_+ =  \sup_{|\psi \otimes \phi\rangle} \,  \<\psi \otimes \phi| (P+Q^\Gamma)| \psi \otimes \phi \> $. However, it is not clear that whether the formula (\ref{WPQM}) provides a decomposable EW, non-decomposable EW or a positive operator. 
In what follows, we provide several examples of optimal non-decomposable EWs for which $W_{\rm M}$ is never non-decomposable, i.e. it is either decomposable EW or a positive operator. 

Finally, we observe that if $W_{\rm M}$ is a mirrored operator to $W$ i.e.
\begin{equation}
    W_{\rm M} = \mu \, \oper_A \otimes \oper_B - W ,
\end{equation}
then 
\begin{equation}
    W^\Gamma_{\rm M} = \mu \, \oper_A \otimes \oper_B - W^\Gamma ,
\end{equation}
is a mirrored operator to $W^\Gamma$ with the same $\mu$. 


\section{Mirroring optimal decomposable entanglement witnesses} \label{SEC-D}

Let us consider a decomposable EW in $\mathbb{C}^n \otimes \mathbb{C}^m$. We start our analysis with extremal decomposable EWs, i.e. $W=|\Psi\>\<\Psi|^\Gamma$ for some entangled state $|\Psi\rangle \in \mathbb{C}^n \otimes \mathbb{C}^m$ \cite{TOPICAL,KYE}. 


\begin{Proposition} \label{Pro-1} If $W = |\Psi\>\<\Psi|^\Gamma$, then the mirrored operator $W_{\rm M}$ is positive semi-definite.
\end{Proposition}

Proof: Let 
\begin{equation}
    |\Psi\rangle = \sum_{k=0} s_k |e_k \otimes f_k \rangle ,
\end{equation}
stand for the Schmidt decomposition of $|\Psi\rangle$ with  $s_0 \geq s_1 \geq s_2 \geq \ldots  \geq 0$.
Note that the maximal eigenvalue of  $|\Psi\>\<\Psi|^\Gamma$ equals $s_0^2$ and it corresponds to the product vector $|e_0 \otimes f_0^*\rangle$. One has therefore
\begin{equation}
    \mu = \sup_{|\psi \otimes \phi \rangle} \<\psi \otimes \phi|\Psi\>\<\Psi|^\Gamma| \psi \otimes \phi\> = s_0^2 ,
\end{equation}
and hence the mirrored operator 
\begin{equation}
    W_{\rm M} = s_0^2 \, \mathbb{I}_n \otimes  \mathbb{I}_m - |\Psi\>\<\Psi|^\Gamma,
\end{equation}
is by construction positive definite. \hfill $\Box$\\

In particular, if $|\Psi\> =|\Psi^+_n\>$ is a maximally entangled state, i.e.    $|\Psi^+_n\> =  \frac{1}{\sqrt{n}} \sum_{k=0}^{n-1} |k \otimes k\rangle$, then $s_0^2=1/n$ and hence
\begin{equation}
    W_{\rm M} = \frac{1}{n} \Big( \mathbb{I}_n \otimes  \mathbb{I}_n - \mathbb{F} \Big) \geq 0 ,
\end{equation}
where $\mathbb{F}$ is a flip (swap) operator defined via
\begin{equation}
    \mathbb{F} = n |\Psi^+_n\>\<\Psi^+_n|^\Gamma = \sum_{i,j=0}^{n-1} |i\rangle \langle j| \otimes |j\rangle \langle i| .
\end{equation}

\begin{remark}
Interestingly, if $W = |\Psi\>\<\Psi|^\Gamma$, then the corresponding SPA is always a separable operator \cite{SPA-1}. Hence, for extremal decomposable EWs both conjectures hold true.  
\end{remark}
Beyond the extremal EWs $W = |\Psi\>\<\Psi|^\Gamma$, we do not have a proof of our conjecture. However, there are several examples supporting it.

\begin{ex} Consider an EW corresponding to the reduction map (\ref{Rn})
\begin{equation}    \label{W-red}
    W = \mathbb{I}_n \otimes  \mathbb{I}_n - n P^+_n ,
\end{equation}
with $P^+_n = |\Psi^+_n\>\<\Psi^+_n|$ being the rank-1 projector onto canonical maximally entangled state. One easily finds $\mu=1$ and hence the mirrored operator
\begin{equation}
    W_{\rm M} =  n P^+_n ,
\end{equation}
is evidently positive definite. Moreover, SPA corresponding to Eq.~(\ref{W-red}) satisfies the SPA conjecture \cite{SPA-1}.
\end{ex}

\begin{ex}

In Ref.~\cite{dar-gni}, the authors have provided a family of decomposable witnesses in 
$\mathbb{C}^3 \otimes \mathbb{C}^3$ mentioned below which violate SPA conjecture:
\begin{equation}  \label{3B}
    W_\gamma=3B_{\gamma}^\Gamma,
\end{equation}
with 
\begin{equation}
B_\gamma=\frac{1-\gamma}{2} P_{10}+\frac{1-\gamma}{2} P_{20}+\gamma P_{11} ,    
\end{equation}
 where $P_{kl}=|\Omega_{kl} \rangle \langle \Omega_{kl}|$ denotes a set of rank-1 projectors with $|\Omega_{kl} \rangle=W_{kl} \otimes \mathbb{I}|\Omega_{00}\rangle$, $W_{kl}$ is a Weyl operator defined by $W_{kl} |i\rangle=w^{k(i-l)}|i-l\rangle$ with $w=e^{2\pi i/3}$. One can express $W_{\gamma}$ explicitly in the following matrix form
 \begin{widetext}
\begin{equation}
    W_{\gamma}=\left(
\begin{array}{ccccccccc}
 1-\gamma  & . & . & . & . & . & . &  \gamma  w & . \\
 . & \gamma  & . & -\frac{1-\gamma}{2}& . & . & . & . & . \\
 . & . & . & . & \gamma  w^* & . & -\frac{1-\gamma}{2} & . & . \\
 . & -\frac{1-\gamma}{2} & . & . & . & . & . & . & \gamma  w^* \\
 . & . & \gamma  w & . & 1-\gamma  & . & . & . & . \\
 . & . & . & . & . & \gamma  & . & -\frac{1-\gamma}{2}  & . \\
 . & . & -\frac{1-\gamma}{2} & . & . & . & \gamma  & . & . \\
 \gamma  w^* & . & . & . & . & -\frac{1-\gamma}{2}  & . & . & . \\
 . & . & . & \gamma  w & . & . & . & . & 1-\gamma  \\
\end{array}
\right) ,
\end{equation}
\end{widetext}
where, to make the formula more transparent, we replaces zeros by dots.
It was proved~\cite{dar-gni} that for each $\gamma \in (0,1)$, $W_{\gamma}$ is an optimal entanglement witness. However, being an optimal EW, it does not satisfy SPA conjecture. Indeed, it turns out \cite{dar-gni} that SPA of $W_\gamma$ defines an entangled state  for some range of $\gamma$, specifically, for $\gamma \in (0.7,1)$. Consider now a mirrored operator 
\begin{equation}
    W_\gamma^\mu=\mu\; \mathbb{I}_3 \otimes \mathbb{I}_3-W_\gamma.
\end{equation}
Clearly $\mu$ depends upon $\gamma$. It turns out that for  $\gamma \geq 0.7$, one has $\mu=\frac{\gamma+1}{2}$. 




Interestingly,  both $W_\gamma$ and $W_\gamma^\mu$ can detect the Bell states. However, we have not been successful to construct a bound entangled state which will be detected by $W_\gamma^\mu$. We have tried all magic simplex states~\cite{magic1,magic2,magic3}, i.e. all possible convex combinations of a complete set of Bell states. Recently, the authors in Refs.~\cite{MagicSimplex1,MagicSimplex2,hiesmayr3} have shown that
with a probability of success of $95\%$, one can solve the separability problem for that huge family of states i.e. for Bell diagonal qutrit states with positive partial transposition. 
However, none of those states can be detected by the witness under our investigation. 
Summing up, our usual methods to find a PPT-entangled state, which is detected by $W_\gamma^\mu$, have failed.

\end{ex}

\section{Mirroring non-decomposable entanglement witnesses}  \label{SEC-ND}

In this Section, we analyze our conjecture for non-decomposable EWs. Again, we do not provide the proof but in what follows, we present several examples supporting the conjecture.



\subsection{EWs from unextendible product bases}
Consider the well known  EW proposed in Refs.~\cite{UPB-1,UPB}: if $|a_k \otimes b_k\rangle \in \mathcal{H}_A \otimes \mathcal{H}_B$ defines an unextendible product basis, i.e. an incomplete orthogonal
product basis in $\mathcal{H}_A \otimes \mathcal{H}_B$ whose complementary subspace contains no product vector, then 
the following operator
\begin{equation}
    W = \sum_k |a_k \otimes b_k\>\<a_k \otimes b_k| - \epsilon_- \oper_{AB} ,
\end{equation}
with
\begin{equation}
    \epsilon_- := \inf_{|a \otimes b\rangle} \, \sum_k |\<a_k \otimes b_k| a \otimes b\>|^2 ,
\end{equation}
defines a non-decomposable EW. One finds for the mirrored operator

\begin{equation}  \label{M-UPB}
    W_{\rm M} = \epsilon_+ \oper_{AB} - \sum_k |a_k \otimes b_k\>\<a_k \otimes b_k|  ,
\end{equation}
with
\begin{equation}
    \epsilon_+ := \sup_{|a \otimes b\rangle} \, \sum_k |\<a_k \otimes b_k| a \otimes b_k\>|^2 = 1.
\end{equation}
Hence, the mirrored operator is a projection onto a subspace orthogonal to ${\rm span} \,\{|a_k \otimes b_k\rangle \}$. After an appropriate normalization, $W_{\rm M}$ defines a PPT state and 
\begin{equation}  \label{M-UPB}
    {\rm Tr}(W W_{\rm M})  = - \epsilon_-  {\rm Tr} W_{\rm M} < 0 , 
\end{equation}
that is, $W_{\rm M}$ is a PPT-entangled operator detected by $W$ \cite{UPB-1}.

\subsection{Choi EW and its generalization in $\mathbb{C}^3 \otimes \mathbb{C}^3$}
\label{33scenario}
As a second example, let us consider a family of EWs in $\mathbb{C}^3 \otimes \mathbb{C}^3$ defined by \cite{Korea-1992}
\begin{eqnarray}
\label{W-abc}
  W[a,b,c] &=& \sum_{i=0}^2 \Big[\, a\, |ii\>\<ii| + b\, |i,i+1\>\<i,i+1| \nonumber\\
  &+& c\, |i,i+2\>\<i,i+2|\, \Big]
   - \sum_{i\neq j=0}^2  |ii\>\<jj|\  ,
\end{eqnarray}
with $a,b,c\geq 0$ satisfying $a+b+c\geq 2$ and if $a \leq 1$, then additionally $bc\geq (1-a)^2$. This family provides a generalization to the well known Choi witness corresponding to $W[1,1,0]$ or $W[1,0,1]$ \cite{Choi-2,Choi-4}. Choi witness was proved to be extremal \cite{Choi-5,Ha-extr} and hence also optimal. Interestingly, being optimal it does not have a spanning property  \cite{S71,S72}. A subclass of $W[a,b,c]$ defined by \cite{Kossak} (cf. also \cite{FilipI}) 
$$ a+b+c=2 \ , \ \ \ a^2+b^2+c^2=2 , $$
was proved to be optimal if and only if $a \in [0,1]$, and  extremal if and only if $a \in (0,1]$. Moreover, $W[a,b,c]$ is decomposable only if $b=c=1$.

\begin{figure}
    \centering
    \includegraphics[width=7.5cm]{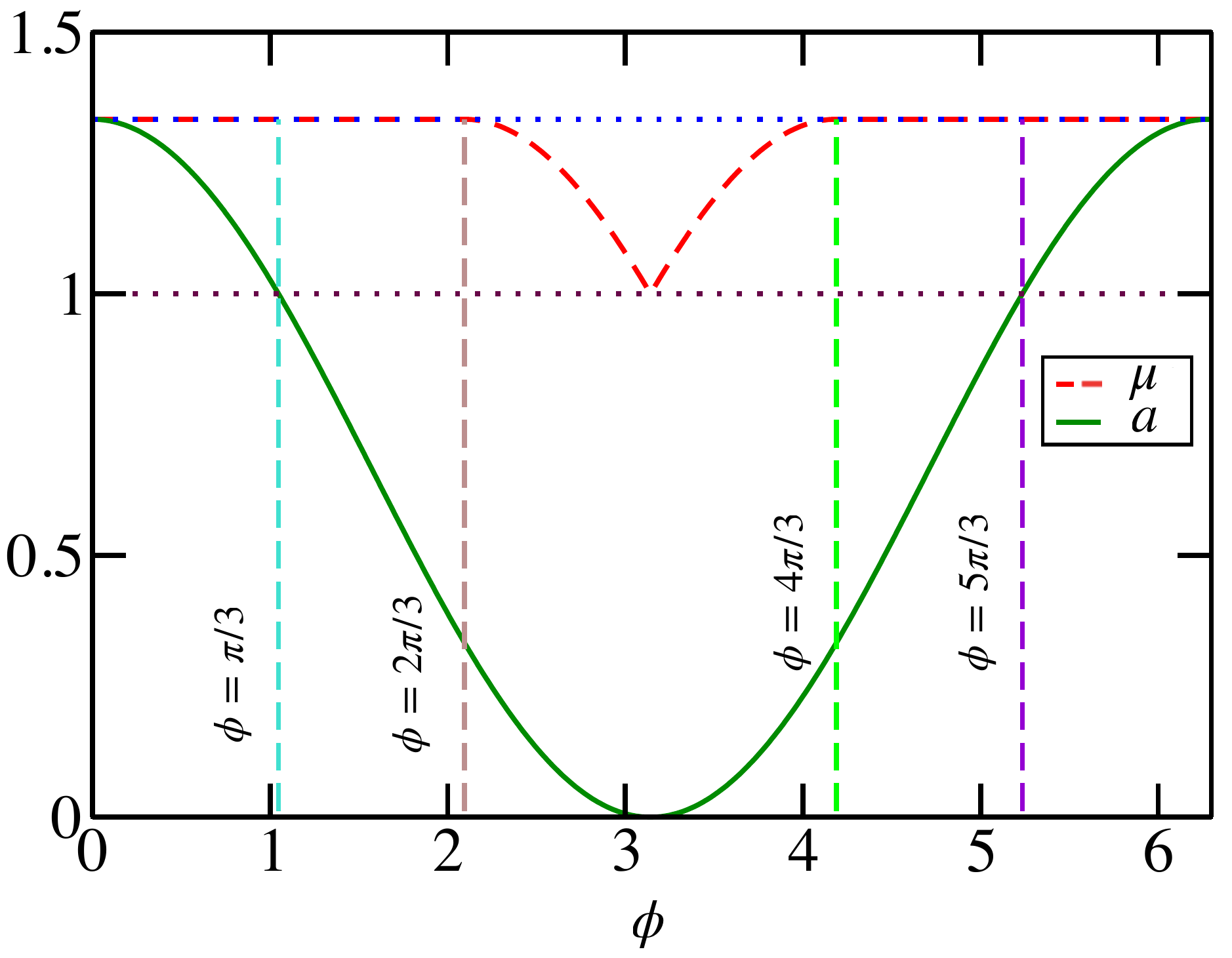}
    \caption{{
    The plot  of $\mu=\mu(\phi)$ and $a=a(\phi)$.}}
    \label{fig33}
\end{figure}

\begin{Proposition} \label{P-abc} The mirrored operator to (\ref{W-abc}) is

\begin{itemize}
    \item positive if $a \in [0,1/3]$,
    \item decomposable EW if $a \in (1/3,4/3]$.
\end{itemize}
\end{Proposition}
Proof: Let us use the following convenient parameterization \cite{Kossak,FilipI}
\begin{eqnarray}
\label{param3}
  a &=& \frac{2}{3} ( 1 + \cos \phi) ,\nonumber\\
  b &=& \frac{1}{3} ( 2 - \cos \phi - \sqrt{3} \sin\phi),\nonumber\\
  c &=& \frac{1}{3} ( 2 - \cos \phi + \sqrt{3} \sin\phi), 
\end{eqnarray}
that is, one has a 1-parameter family of EWs $W(\phi)$ for $\phi \in [0,2\pi)$. $W(\pi/3)$ and $W(5\pi/3)$ correspond to a pair of Choi witnesses and $W(\pi)$ corresponds to EW defined via the reduction map $R_3$. $W(\phi)$ is optimal iff $\phi \in [\pi/3,5\pi/3]$. One easily finds for $\mu(\phi)$ (cf. the Figure \ref{fig33})

\begin{equation}
    \mu(\phi) = \left\{ \begin{array}{ll} 4/3   & ; \ \ \phi \in [0,2\pi/3] \cup [4 \pi/3,2\pi) \\
    c(\phi)    & ; \ \ \phi \in [2\pi/3,\pi] \\
     b(\phi)    & ; \ \ \phi \in [\pi,4 \pi/3] \end{array}  \right. 
\end{equation}
such that the mirrored operator
\begin{equation}
    W_{\rm M}(\phi) = \mu(\phi) \oper_3 \otimes \oper_3 - W(\phi) ,
\end{equation}
is block-positive. Note, that $b(\pi)=c(\pi)=1$. Now, for $\phi \in [0,2\pi/3] \cup [4 \pi/3,2\pi)$ one has
\begin{equation}
\label{naam1}
    W_{\rm M}(\phi) = \frac 43 \, \oper_3 \otimes \oper_3 - W(\phi)  = 3 \left(\frac 43 -a \right) P^+_3 + B^\Gamma(\phi) , 
\end{equation}
with

\begin{eqnarray}
    B(\phi) &=& \left(\frac 43 - b\right) \sum_{i=0}^2 |i\>\<i|\otimes |i+1\>\<i+1|
     \nonumber\\
   &+&  \left(\frac 43 - c\right) \sum_{i=0}^2 |i\>\<i|\otimes |i+2\>\<i+2| \nonumber\\
  &+&  \left(a-\frac 13\right) \sum_{i\neq j=0}^2 |i\>\<j|\otimes |j\>\<i| .
\end{eqnarray}

  Indeed, for $\phi \in [0,2\pi/3] \cup [4 \pi/3,2\pi)$ one has $a,b,c \leq 4/3$. Therefore, the first part of Eq.~\eqref{naam1} is positive. We now will show that $B(\phi)\geq 0$. Note that the positivity of $B(\phi)$ is equivalent to positivity of the following $2\times 2$ submatrix

\begin{equation}
    \left(
\begin{array}{cc} 4/3 - b & a - 1/3 \\ a - 1/3 & 4/3 - c \end{array} \right) .
\end{equation}
Simple calculation shows that the determinant  of this submatrix equals to $0$, 
which proves that $B(\phi) \geq 0$ and hence $W_{\rm M}(a,b,c)$ is decomposable. 

Now, if $\phi \in (2\pi/3,\pi]$, we can express $W_{\rm M}$ in the following form
\begin{eqnarray}
    W_{\rm M}(\phi) &=& (c-a) \sum_{i=0}^2 |i\>\<i|\otimes |i\>\<i| \nonumber\\
    &+& \left(c - b\right) \sum_{i=0}^2 |i\>\<i|\otimes |i+1\>\<i+1| \nonumber\\
    &+& \sum_{i\neq j=0}^2 |i\>\<j|\otimes |i\>\<j|,
\end{eqnarray}
where
$$  c-a = \frac 13(\sqrt{3} \sin \phi - 3 \cos\phi) \geq 0, \ \ \ \ c-b = \frac{2}{\sqrt{3}}\, \sin\phi \geq 0. $$
This proves that $W_{\rm M}(\phi)\geq 0$ in $\phi \in (2\pi/3,\pi]$. Similar analysis shows that $W_{\rm M}(\phi) \geq 0$
for $\phi \in [\pi,4 \pi/3]$. We summarize our finding in Table~\ref{table1}.   \hfill $\Box$

\begin{remark} Note that if $W_{\rm M}(\phi) \geq 0$, i.e. $\phi \in [2\pi/3,4\pi/3]$, then $\tr[W_{\rm M}(\phi)W(\phi)]<0 $, i.e. 
$ \rho = W_{\rm M}(\phi)/{\rm Tr}W_{\rm M}(\phi)$ defines a PPT entangled state. 
Indeed,  one has
\begin{equation}
    \tr[W_{\rm M}(\phi)W(\phi)] = \mu(\phi) \tr W(\phi) - \tr [W(\phi)W(\phi)] . 
\end{equation}
Note, that $\tr W(\phi) = 3(a+b+c)=6$ and $\tr [W(\phi)W(\phi)] = 12$, and hence
\begin{equation}
     \tr[W_{\rm M}(\phi)W(\phi)] = 6(\mu(\phi)-2)  
\end{equation}
is always negative due to $\mu(\phi) \leq 4/3$.   
\end{remark}


\begin{center}
\begin{table}[h]
\begin{tabular}{ |c|c|c|c|c|}
\hline
 \multicolumn{1}{|c|}{$\phi$} & \multicolumn{1}{|c|}{Optimality of $W$} & \multicolumn{1}{|c|}{ND of $W$} & \multicolumn{1}{c|}{$\mu$}  & \multicolumn{1}{c|}{${W_{\rm M}}$}  \\\hline\hline
 $[\pi/3,2 \pi/3]$ & \cmark & \cmark &  $4/3$ & D \\\hline
  $[2 \pi/3, \pi)$ & \cmark  & \cmark  & $c(\phi)$  & PO   \\\hline
   $\pi$ & \cmark & \xmark & $4/3$ & PO \\\hline
  $(\pi, 4 \pi/3] $ & \cmark & \cmark & $b(\phi)$  & PO \\\hline
  $[4 \pi/3, 5 \pi/3]$ & \cmark & \cmark &  $4/3$ & D \\\hline

\end{tabular}
\caption{The regions of decomposability (D) or non-decomposability (ND) or positive operator (PO) for different values of $\phi$ in the scenario of entanglement witnesses $W(\phi)$ and their corresponding mirrored ones $W_{\rm M}(\phi)$ in $\mathbb{C}^3 \otimes \mathbb{C}^3$. 
}
\label{table1}
\end{table}
\end{center}

\subsection{Mirrored pairs in $\mathbb{C}^4 \otimes \mathbb{C}^4$}

Let us consider now a family of EWs being a generalization of a family $W[a,b,c]$ in $\mathbb{C}^4 \otimes \mathbb{C}^4$ \cite{PRA-2022}
\begin{eqnarray}
\label{W-abcd}
  W[a,b,c,d] &=& \sum_{i=0}^3 \Big[\, a\, |ii\>\<ii| + b\, |i,i+1\>\<i,i+1| \nonumber\\
  &+& c\, |i,i+2\>\<i,i+2| +  d\, |i,i+3\>\<i,i+3|\, \Big] \nonumber\\
   &-& \sum_{i\neq j=0}^3  |ii\>\<jj|, 
\end{eqnarray}
with $a,b,c,d\geq 0$ satisfying 
\begin{eqnarray}  
  a+b+c+d &=& a^2+b^2+c^2+d^2 = 3, \label{II-1}\\
  ac+bd &=& 1 , \ \ \ \  (a+c)(b+d) =  2 .
\end{eqnarray}
There are two solutions to the above set of equations \cite{PRA-2022}:  class I is characterized by (\ref{II-1}) together with
\begin{equation}\label{I}
  a+c=2 \ , \ \ b+d=1 \ ,
\end{equation}
whereas class II is characterized by (\ref{II-1}) together with
\begin{equation}\label{II}
  a+c=1 \ , \ \ b+d=2 \ .
\end{equation} 
Interestingly, it is shown \cite{PRA-2022} that EWs from class I are not optimal, whereas those from class II are optimal.

\begin{ex} Consider a Choi-like EWs $W[1,1,1,0]$. Contrary to $W[1,1,0]$ in $\mathbb{C}^3 \otimes \mathbb{C}^3$,  it is not optimal. One easily finds the corresponding mirrored operator
\begin{equation}
    W_{\rm M}[1,1,1,0] = \frac 43\, \oper_4 \otimes \oper_4 - W[1,1,1,0] .
\end{equation}
It turns out that the mirrored operator $ W_{\rm M}[1,1,1,0]$ defines a non-decomposable EW. Indeed, by considering the following (unnormalized) state
\begin{eqnarray}
\label{}
  \rho_x &=& \sum_{i=0}^3 \Big[\, 3\, |ii\>\<ii| + x\, |i,i+1\>\<i,i+1| + \, |i,i+2\>\<i,i+2| \nonumber\\
  &+&  \frac 1x\, |i,i+3\>\<i,i+3|\, \Big]
   - \sum_{i\neq j=0}^3  |ii\>\<jj|\  ,
\end{eqnarray}
where $x > 0$, it is easy to check that $\rho_x$ is PPT.  Hence, we obtain
\begin{equation}
    \tr(W_{\rm M}[1,1,1,0]\, \rho_x) = \frac{4}{3x}(x^2-5x +4). 
\end{equation}
Clearly, $ \tr(W_{\rm M}[1,1,1,0]\, \rho_x) < 0$ if and only if $x \in (1,4)$. This proves that $W_{\rm M}[1,1,1,0]$ detects a PPT-entangled state $\rho_x$ and therefore, it is a non-decomposable EW.  It is evident that in this case we have a pair of mirrored non-decomposable EWs $( W[1,1,1,0], W_{\rm M}[1,1,1,0])$. 
This example shows that if one relaxes the requirement of optimality, then the mirrored operator might be non-decomposable EW as well.  
\end{ex}

Similar to the witnesses $W[a,b,c]$ in $\mathbb{C}^3 \otimes \mathbb{C}^3$,  $W[a,b,c,d]$ can be parameterized as follows:
\begin{eqnarray}
\mbox{ class I}:~~~a &=& \frac{1}{2} (2-\sin\theta)  \ , \ b = \frac{1}{2} (1+\cos\theta) \ , \nonumber\\
\ c &=& 2-a \ , ~~~~~~~~~~ d = 1- b \ ,
\label{eq:theta_class1}
\end{eqnarray}
and
\begin{eqnarray}
\mbox{ class II}:~~~a &=& \frac{1}{2} (1+\cos\theta) \ , \ b = \frac{1}{2} (2-\sin\theta) \ , \nonumber\\
\ c &=& 1-a\ , ~~~~~~~~~~ d = 2-b \ ,
\label{eq:theta_class2}
\end{eqnarray}
with $\theta \in[0,\pi]$. We use the notations $W_I(\theta)$ and $W_{II}(\theta)$ for  $W[a,b,c,d]$ in the first and second classes, respectively.
In particular, $W_I(0) = W[1,1,1,0]$ and 
$W_I(\pi) = W[1,0,1,1]$ are Choi-like EWs, $W_I(\pi/2) = W[1/2,1/2,3/2,1/2]$ is the only decomposable EW in the class I. 
Similarly, $ W_{II}(\pi) = W[0,1,1,1]$ corresponds to the reduction map $R_4$, whereas    $W_{II}(0) = W[1,1,0,1]$  is the second decomposable EW in the class II (cf. \cite{PRA-2022}). 

For $\theta \in (0,\pi)$, the class II consists of non-decomposable EWs. One finds for the mirrored operators
\begin{equation}
    W^{\rm M}_{II}(\theta) = \mu(\theta) \oper_4 \otimes \oper_4 - W_{II}(\theta) , 
\end{equation}
with
\begin{equation}
    \mu(\theta) = \left\{ \begin{array}{ll} 3/2  & ; \ \ \theta \in (0,\pi/2) \\
    d(\theta)    & ; \ \ \theta \in (\pi/2,\pi) \end{array}  \right. 
\end{equation}

\begin{Proposition} \label{Pa} The mirrored operator to $W_{II}(\theta)$ is

\begin{itemize}
    \item decomposable EW if $\theta \in (0,\pi/2)$,
     \item positive if $\theta \in (\pi/2,\pi)$.
\end{itemize}
\end{Proposition}
The proof is very similar to that of Proposition \ref{P-abc} (cf. Appendix \ref{appA}). 

\begin{remark} Note that if $W^{\rm M}_{II}(\theta) \geq 0$, i.e. $\theta \in (\pi/2,\pi)$, then $\tr[W_{II}(\theta)W^{\rm M}_{II}(\theta)]<0 $, that is, $ \rho = W^{\rm M}_{II}(\theta)/{\rm Tr}W^{\rm M}_{II}(\theta)$ defines a PPT enatgled state. Indeed, one has
\begin{eqnarray}
    \tr[W^{\rm M}_{II}(\theta)W_{II}(\theta)] = \mu(\theta) \tr W_{II}(\theta) - \tr [W_{II}(\theta)W_{II}(\theta)] . \nonumber\\
\end{eqnarray}
Note, that $\tr W_{II}(\theta) = 4(a+b+c+d)=12$ and $\tr [W_{II}(\theta)W_{II}(\theta)] = 24$, and hence
\begin{equation}
    \tr[W_{II}(\theta)W^{\rm M}_{II}(\theta)] = 12(\mu(\theta)-2)  
\end{equation}
is always negative due to $\mu(\theta) \leq 3/2$.   
\end{remark}

Now, the class I contains non-optimal EWs. In Ref.~\cite{PRA-2022} by following the paper \cite{Lew}, an optimization procedure was performed leading to an optimal EW defined via
\begin{equation}
\label{opti_c1}
    \tilde{W}_I(\theta) = W_I(\theta) - 2 P,
\end{equation}
where $P=|\Psi\>\<\Psi|$  is a rank-1 projector onto the maximally entangled state in $\cc^4 \otimes \cc^4$ with   $|\Psi\> = \frac 12 \sum_{j=0}^3 (-1)^{j} |j \otimes j\>$. 

\begin{Proposition} \label{Pb} The mirrored operator to $\tilde{W}_{I}(\theta)$ for $\theta \in [0,\pi] - \{\pi/2\}$
\begin{equation}
    \tilde{W}^{\rm M}_{I}(\theta) = \frac 32 \oper_4 \otimes \oper_4  - \tilde{W}_{I}(\theta) , 
\end{equation}
is a decomposable EW.
\end{Proposition}
The proof is very similar to that of Proposition \ref{P-abc} (cf. Appendix \ref{AppB}).

\subsection{A class of non-decomposable  Breuer-Hall maps}

In Refs.~\cite{Breuer1, Hall1}, Breuer and Hall have generalized the reduction map by the following class of positive maps 
$\Phi : M_{2n}(\mathbb{C}) \rightarrow M_{2n}(\mathbb{C})$ such that
\begin{equation}
\label{bh1}
\Phi_{\rm BH}(X)=R_{2n}(X)-UX^TU^\dagger,
\end{equation}
where $U$ is an arbitrary antisymmetric unitary matrix in $ M_{2n}(\mathbb{C})$.   It was shown that this map is non-decomposable~\cite{Breuer1, Hall1} and optimal~\cite{Breuer1}, even nd-optimal~\cite{maciej00,Breuer1}. The mirrored positive map
\begin{equation}
\label{bh1}
\Phi_{\rm BH}^{\rm M}(X)= \oper_{2n} \tr X - \Phi_{\rm BH}(X) = X + UX^TU^\dagger,
\end{equation}
which is evidently decomposable being a sum of an identity and completely co-positive 
map $UX^TU^\dagger$.

\subsection{A class of non-decomposable maps in $M_n(\mathbb{C}) \otimes M_n(\mathbb{C})$} 

 Let 
 $\varepsilon: M_n(\mathbb{C}) \to M_n(\mathbb{C})$ be the canonical projection of $M_n(\mathbb{C})$ to the diagonal part
\begin{equation}
\varepsilon(X)=\sum_{i=0}^{n-1} \<i|X| {i}\> |i\>\< i| .
\end{equation}
Let $S$ be a  permutation  defined by
\begin{equation}
S |i\>=|i+1\>,  \ \ \ \ (\mbox{mod}~n),
\end{equation}
for $i=0,1,\ldots, n-1$. The following  maps $\tau_{n,k} : M_n(\mathbb{C}) \to M_n(\mathbb{C})$
\begin{equation}  \label{!}
\tau_{n,k}(X)=(n-k) \varepsilon(X)+\sum_{i=1}^{k} \varepsilon\big(S^i X S^{\dagger i}\big)-X ,
\end{equation}
for $k=0,1,\ldots,n-1$, were proved to be  positive and non-decomposable if $k<n-1$ \cite{TT,Osaka1,Osaka2,Ando,RIMS,yamagami}. Actually, $\tau_{n,n-1} = R_n$ (reduction map), and $\tau_{3,1}$ is a Choi map in $M_3(\mathbb{C})$. 

Now, the mirrored map is defined via
\begin{equation}
    \tau^{\rm M}_{n,k}(X) = \mu_{n,k}\, \oper_n \tr X - \tau_{n,k}(X) .
\end{equation}
Let us denote the greatest common divisor of $n$ and $k$ by ${\rm gcd}(n,k)$. In a recent paper \cite{LAA}, the authors have shown that if 
${\rm gcd}(n,k)=1$, then $\tau_{n,k}$ is optimal. In particular, ${\rm gcd}(n,1)=1$ and we find that
\begin{equation}
    \mu_{n,1} = \left\{ \begin{array}{ll} \frac 43 \ &; \ n=3, \\ n-2 \ &; \  n \geq 4  . \end{array} \right.
\end{equation}
Similarly, for odd $n$, one has ${\rm gcd}(n,n-2)=1$ and  we obtain
\begin{equation}
    \mu_{n,n-2} = \left\{ \begin{array}{ll} \frac 43 \ &; \ n=3, \\ \frac  32 \ &; \  n \geq  4 . \end{array} \right.
\end{equation}

\begin{Proposition} The mirrored maps $ \tau^{\rm M}_{n,1}$ and  $\tau^{\rm M}_{n,n-2}$  are decomposable.
\end{Proposition}
Proof: Due to Choi-Jamio{\l}kowski isomorphism, the EW corresponding to the map $\tau_{n,k}$ can be expressed as
\begin{eqnarray}
    W_{n,k} &=& \sum_{i,j=0}^{n-1} |i\>\<j| \otimes \tau_{n,k}(|i\>\<j|) \nonumber\\
    &=& \sum_{i=0}^{n-1}\Big[ (n-k-1)|ii\>\<ii|+ \sum_{\ell=1}^k |i,i+\ell\>\<i,i+\ell| \Big] \nonumber\\
    &-& \sum_{i\neq j} |ii\>\<jj|. 
\end{eqnarray}
Therefore,  for $n \geq 4$, one can easily rite the corresponding mirrored operator
\begin{eqnarray}
W^{\rm M}_{{n,1}} &=& \sum_{i=0}^{n-1} \Big[ (n-3) |i,i+1\rangle \langle i, i+1| \nonumber\\ &+& (n-2) \sum_{\ell=2}^{n-1} |i,i+\ell\>\<i,i+\ell| \Big] 
+ \sum_{i \neq j=0}^{n-1} |ii \rangle \langle jj |.   \nonumber\\
\end{eqnarray}
Note, that $W^{\rm M}_{{n,1}} = B_{n,1}^\Gamma$, with $B_{n,1}^\Gamma \geq 0$. Actually, positivity of $B_{n,1}$ is equivalent to the positivity of the following $2\times 2$ submatrix
\begin{equation}
    \left(
\begin{array}{cc}
 n-3 & 1  \\
 1 & n-2 
\end{array}
\right),
\end{equation}
which is evidently positive for $n\geq 4$. This implies that  $W^{\rm M}_{\tau_{n,1}}$ is decomposable for $n \geq 4$.

Now, for the odd $n$, $\tau_{n,n-2}$ produces again the Choi map for $n=3$ that we discussed above. For $n>3$, the corresponding mirrored EW for $\mu=3/2$ can be expressed as follows

\begin{eqnarray}
    W^M_{{n,n-2}} &=& \frac{1}{2} \sum_{i=0}^{n-1} \Big[ \sum_{\ell=0}^{n-2}  |i,i+\ell\rangle \langle i,i+\ell| \nonumber\\ &+& 3|i,i+n-1\>\<i,i+n-1| \Big] + \sum_{i \neq j=0}^{n-1} |ii \rangle \langle jj |. \nonumber\\
\end{eqnarray}
Moreover, $W^M_{\tau_{n,n-2}}$ can be written as $W^M_{{n,n-2}}=A_{n,n-2}+B_{n,n-2}^\Gamma$, where
\begin{equation}
    A_{n,n-2} =\frac{1}{2} \sum_{i \neq j=0}^{n-1} \Big[|ii\rangle \langle ii|+ |ii\rangle \langle jj| \Big] >0,
\end{equation}
and
\begin{eqnarray}
    B_{n,n-2} &=& \frac{1}{2} \Big[ \sum_{i=0}^{n-1} \Big( \sum_{\ell=1}^{n-2} |i,i+\ell\rangle \langle i, i+\ell| \nonumber\\ &+& 3|i,i+n-1\>\<i,i+n-1| \Big) + \sum_{i \neq j=0}^{n-1} |ij \rangle \langle ji| \Big] . \nonumber\\
\end{eqnarray}
It is easy to see that $B_{n,n-2}$ is positive as the positivity of $B_{n,n-2}$ is equivalent to the positivity of the following $2\times 2$ submatrix
\begin{equation}
    \left(
\begin{array}{cc} 1 & 1 \\ 1 & 3 \end{array} \right) .
\end{equation}
This shows  that  $W^M_{{n,n-2}}$ is decomposable for $n \geq 4$.  \hfill $\Box$


\subsection{A family of optimal non-decomposable witnesses in $\mathbb{C}^3 \otimes \mathbb{C}^3$ whose SPA is not separable}

In this section, we consider a family of indecomposable entanglement witnesses proposed by Ha and Kye in Ref.~\cite{hakyend} whose SPAs are not separable. For non-negative real numbers $a$, $b$, $c$ and $-\pi \leq \theta \leq \pi$, the form of the self-adjoint block matrix in $\mathbb{C}^3 \otimes \mathbb{C}^3$ is given by
\begin{eqnarray}   \label{W-theta}
    W[a,b,c;\theta]=\left(
\begin{array}{ccc|ccc|ccc}
 a & . & . & . & -e^{i \theta } & . & . & . & -e^{-i \theta } \\
 . & b & . & . & . & . & . & . & . \\
 . & . & c & . & . & . & . & . & . \\  \hline
 . & . & . & c & . & . & . & . & . \\
 -e^{-i \theta } & . & . & . & a & . & . & . & -e^{i \theta } \\
 . & . & . & . & . & b & . & . & . \\  \hline
 . & . & . & . & . & . & b & . & . \\
 . & . & . & . & . & . & . & c & . \\
 -e^{i \theta } & . & . & . & -e^{-i \theta } & . & . & . & a \\
\end{array}
\right). \nonumber\\
\end{eqnarray}
Let $p_\theta=\max \{q_{(\theta-\frac{2}{3} \pi)},q_\theta,q_{(\theta+\frac{2}{3} \pi)} \}$, where $q_\theta=e^{i \theta}+e^{-i\theta}$. One has $1 \leq p_\theta \leq 2$. Now,  $W[a,b,c;\theta] \geq 0$  iff $a \geq p_\theta$ and $W[a,b,c;\theta]$ is  is  block-positive iff the following conditions hold
\begin{equation}
    1) \ \ a+b+c \geq p_\theta , \ \   \ \ 2) \ \ \mbox{if} \  a\leq 1 , \ \ \mbox{then} \ \  bc \geq (1-a)^2.
\end{equation}
The authors of \cite{hakyend} analyzed two classes of EWs:
\begin{equation}
    \label{C1}
               2-p_\theta \leq a<1, \ \ \ a+b+c=p_\theta, \ \ \ bc=(1-a)^2,
    \end{equation}
and 
 \begin{equation}  \label{C2}
        1 \leq a<p_\theta,\ \ \ a+b+c=p_\theta,\ \ \ bc=0,
    \end{equation}
For (\ref{C1}) one has  $4/3 \leq p_\theta < 1+1/\sqrt{2}$ and for (\ref{C2}) one has $1+1/\sqrt{2} \leq p_\theta < 2$. Both classes consist of non-decomposable EWs. Moreover, class (\ref{C1}) has the bi-spanning property, whereas class (\ref{C2}) has the co-spanning property \cite{hakyend}. For the first class let us consider $p_\theta=4/3$, $a=2-p_\theta=2/3$, and $b=c=p_\theta-1=1/3$. Then the corresponding mirrored EW is $W'=\mu \mathbb{I}_3 \otimes \mathbb{I}_3-W[a,b,c;\theta]$, where $\mu=1.097$. Let us observe that one can easily construct a PPT state 
\begin{eqnarray}
    \rho_x &=& \sum_{i=0}^2 \Big[ x |i,i\> \<i,i|  +  |i+1,i+1\> \<i+1,i+1| \nonumber\\ &+& |i+2\> \<i+2| + e^{i\theta} |i,i\> \<i+1,i+1| \nonumber\\ &+& e^{-i\theta}|i,i\> \<i+2,i+2|  \Big] ,
\end{eqnarray}
which is PPT if and only if 
\begin{equation}
    \left( \begin{array}{ccc} x & e^{i\theta} & e^{-i\theta} \\ e^{-i\theta} & x & e^{i\theta} \\ e^{i\theta} & e^{-i\theta} & x \end{array} \right) \geq 0 ,
\end{equation}
that is, $x \geq 1$ and $x^3 + 2 \cos(3\theta) - 3 x \geq 0$ which  for $p_\theta = 4/3$ implies $x \geq 1.896$. One finds
\begin{equation}
    {\rm Tr}(W[2/3,1/3,1/3,\theta]\, \rho_x) = 2(x-2) < 0 ,
\end{equation}
for $x \in [1.896,2)$ which proves that $W[2/3,1/3,1/3,\theta]$ is non-decomposable. Consider now the following state 
\begin{eqnarray}
    \rho_y &=&  \sum_{i=0}^2 \Big[ y |i,i\> \<i,i|  +  |i+1,i+1\> \<i+1,i+1| \nonumber\\ &+& |i+2\> \<i+2| -   e^{i\theta} |i,i\> \<i+1,i+1| \nonumber\\ 
    &-& e^{-i\theta}|i,i\> \<i+2,i+2|  \Big] ,
\end{eqnarray}
which is PPT if and only if 
\begin{equation}
    \left( \begin{array}{ccc} y & -e^{i\theta} & -e^{-i\theta} \\ -e^{-i\theta} & y & -e^{i\theta} \\ -e^{i\theta} & -e^{-i\theta} & y \end{array} \right) \geq 0 ,
\end{equation}
that is, $y \geq 1$ and $xy^3 - 2 \cos(3\theta) - 3 y \geq 0$ which,  for $p_\theta = 4/3$, implies $y \geq 3/2$. One finds $
{\rm Tr}((\mu \mathbb{I}_3 \otimes \mathbb{I}_3 - W[2/3,1/3,1/3,\theta])\, \rho_y) > 0.5185$ for $y \geq 3/2$. Clearly, it does not prove that the witness is decomposable. However, both the witness and the state $\rho_y$ have the same symmetry and it is natural to expect that a proper PPT state which detects non-decomposability of the witness belongs to the above family of states. Additionally, we performed numerical analysis and  used all bound entangled states found in the magic simplex, i.e. Bell diagonal states, based on a grid approach (140 000 states)~\cite{MagicSimplex1} and based on a representative random sample which fills the volume of the bound entangled states within the magic simplex~~\cite{MagicSimplex2}. Note that the classification of bound entangled states in the magic simplex is obtained with a very high success rate of $5\%$. Obviously, the given witness must not be sensitive to Bell diagonal states. Therefore, we used the sequentially constrained Monte Carlo method introduced in Ref.~\cite{EnglertBound} to sample from a random set of $100\; 000$ states in this case $8007$ bound entangled states detected by the realignment criterion. Each state was also optimized over local unitaries with the convenient composite parametrization introduced in Ref.~\cite{CPHiesmayr}. None of these bound entangled states detects non-decomposability of $W'$.
    
For the second class (\ref{C2}) we consider  $p_\theta= 1+1/\sqrt{2}$ and hence $a=1, b=p_\theta -1$, $c=0$, or $a=1$, $b=0$, $c=p_\theta-1$. Hence the corresponding mirrored EW is $W''=\mu\; \mathbb{I}_3 \otimes \mathbb{I}_3-W[a,b,c;\theta]$, where $\mu'=1.21473$.   A similar analysis as for the class (\ref{C1}) supports our conjecture. Numerical analysis shows that  no bound entangled state is detected by this witness.


\section{Conclusions }   \label{SEC-C}

 As entanglement is generally a useful resource in quantum information theory, its verification has both fundamental and practical significance. It is, however, generally challenging as its computational complexity lies in the NP-hard class. The problem is also {connected to} a long--standing open question about the classification of positive maps. In this paper, we have approached the problem by exploiting the convexity of separable states and EWs {since optimal EWs define the set of separable states and non-decomposable EWs classify the set of PPT-entangled and free entangled states. PPT-entangled states are of particular interest related to numerous open questions in quantum information theory such as Bell inequalities, activation of entanglement, channel capacities, invariance under Lorentz boosts~\cite{HiesmayrLorentzboost}, etc. Only limited knowledge is known about the structure of the PPT-entangled states in the Hilbert space and also only for low dimensions. We provide here a different view via the structure of connected entanglement witnesses.
 
Along with the convexity of quantum states and EWs, there has been the so-called SPA conjecture that addresses SPAs to optimal EWs are separable states. Counterexamples, however, exist. 
In our work, we have considered the framework of mirrored EW i.e. `twin' of an EW such that both the
EWs can detect the entangled states by realizing a single observable.
A trade-off relation is observed between the EW and its mirrored ones, which we have  presented} as a conjecture in our paper.  Our conjecture states that mirrored operators to optimal EWs are either quantum states or decomposable EWs, { hence} cannot detect the PPT-entangled states.
 In other words, there does not exist a mirrored pair of non-decomposable EWs such that at least one of them is optimal.  Consequently, if our conjecture holds generally, then there is a relation between optimality of an EW and decomposability.

We have proved that mirrored EWs to extremal decomposable witnesses are positive semi-definite i.e., quantum states. In fact, for the extremal decomposable EWs, both our conjecture and the SPA one hold true.
For non-decomposable EWs, several examples that support our conjecture are presented. In particular, those examples that disproved the SPA conjecture are considered: for all of the cases, their mirrored operators cannot detect PPT-entangled states. Let us reiterate that the assumption of optimality is essential: otherwise, one can immediately find examples of non-optimal and non-decomposable EWs such that their mirrored operators are also non-decomposable EWs. 

We believe that that our analysis unfolds a hidden structure of the set of entanglement witnesses, which brings us closer to the understanding of the separability problem.   

\section*{Acknowledgements}
AB and DC were supported by the Polish National Science Centre project No. 2018/30/A/ST2/00837. JB was supported by National Research Foundation of Korea (NRF-2021R1A2C2006309, NRF-2020K2A9A2A15000061), Institute of Information \& communications Technology Planning \& Evaluation (IITP) grant (the ITRC Program/IITP-2021-2018-0-01402).  BCH acknowledges gratefully that this research was funded in whole, or in part, by the  Austrian Science Fund (FWF) project P36102. For the purpose of open access, the author has applied a CC BY public copyright licence to any Author Accepted Manuscript version arising from this submission.

\appendix

\section{Proof of Proposition \ref{Pa}}
\label{appA}
For $\theta \in (0,\pi/2)$, the corresponding mirrored operator of $W_{II}(\theta)$ can be written as
\begin{equation}
\label{c2c1}
    W^{\rm M}_{II}(\theta) = \frac 32 \, \oper_4 \otimes \oper_4 - W_{II}(\theta)  = 4 \left(\frac 32 -a \right) P^+_4 + B_1^\Gamma(\theta) , 
\end{equation}
with
\begin{eqnarray}
    B_1(\theta) &=& \left(\frac 32 - b\right) \sum_{i=0}^3 |i\>\<i|\otimes |i+1\>\<i+1| \nonumber\\ &+& \left(\frac 32 - c\right) \sum_{i=0}^3 |i\>\<i|\otimes |i+2\>\<i+2| \nonumber\\ &+& \left(\frac 32 - d\right) \sum_{i=0}^3 |i\>\<i|\otimes |i+3\>\<i+3| \nonumber\\
   &+& \left(a-\frac 12\right) \sum_{i\neq j=0}^3 |i\>\<j|\otimes |j\>\<i| .
\end{eqnarray}
 Clearly, for $\theta \in (0,\pi/2)$, one has $\frac 12 \leq a,b \leq 1,~0 \leq c \leq \frac 12,~1 \leq d \leq \frac 32$. Hence, the first part of Eq.~\eqref{c2c1} is positive. Now we need to show that $B_1(\theta)\geq 0$. Note that the positivity of $B_1(\theta)$ is equivalent to the positivity of the following two $2\times 2$ submatrices
\begin{eqnarray}
  A_1=  \left(
\begin{array}{cc} \frac 32 - b & a - \frac 12 \\ a - \frac 12 & \frac 32 - d \end{array} \right)~~\mbox{and}~~
A_2=\left(
\begin{array}{cc} \frac 32 - c & a - \frac 12 \\ a - \frac 12 & \frac 32 - c \end{array} \right). \nonumber\\
\end{eqnarray}
Simple calculation shows that $\det[A_1]=0$ and  $\det[A_2]=1+\cos \theta=2a \geq 0$ in $\theta \in (0,\pi/2)$.
This proves that $B_1(\theta) \geq 0$ and hence $W_{II}^{\rm M}(\theta)$ is decomposable.

Now, for $\theta \in (\pi/2,\pi)$, one can express $W^{\rm M}_{II}$ in the following way
\begin{eqnarray}
    W^{\rm M}_{II}(\theta) &=& (d-a) \sum_{i=0}^3 |i\>\<i|\otimes |i\>\<i| 
   \nonumber\\ &+& \left(d - b\right) \sum_{i=0}^3 |i\>\<i|\otimes |i+1\>\<i+1|
   \nonumber\\ &+& \left(d - c\right) \sum_{i=0}^3 |i\>\<i|\otimes |i+2\>\<i+2|
   \nonumber\\ &+&  \sum_{i\neq j=0}^3 |i\>\<j|\otimes |i\>\<j| .
\end{eqnarray}
where
\begin{eqnarray}
    d-a &=& \frac{1}{2} (\sin \theta -\cos \theta +1) \geq 0, ~~ d-b = \sin\theta \geq 0, \nonumber\\ 
    d-b &=& \frac{1}{2} (\sin \theta +\cos \theta +1) \geq 0.
\end{eqnarray}
This proves that $W^{\rm M}_{II}(\theta)\geq 0$ in $\theta \in (\pi/2,\pi)$.


\section{Proof of Proposition \ref{Pb}}
\label{AppB}
 The mirrored operator corresponding  to the optimal EW $\tilde{W}_{I}(\theta)$ for $\theta \in [0,\pi] - \{\pi/2\}$ can be expressed as
\begin{equation}
    \tilde{W}^{\rm M}_{I}(\theta) = \frac 32 \oper_4 \otimes \oper_4  - \tilde{W}_{I}(\theta) =4 \left(2-a \right) P^+_4 + B_2^\Gamma(\theta),
\end{equation}
with
\begin{eqnarray}
\label{c2c2}
    B_2(\theta) &=& \left(\frac 32 - b\right) \sum_{i=0}^3 |i\>\<i|\otimes |i+1\>\<i+1| \nonumber\\
    &+& \left(\frac 32 - c\right) \sum_{i=0}^3 |i\>\<i|\otimes |i+2\>\<i+2| 
    \nonumber\\
    &+& \left(\frac 32 - d\right) \sum_{i=0}^3 |i\>\<i|\otimes |i+3\>\<i+3| \nonumber\\
   &+& \left(a-\frac 32\right) \sum_{\substack{i=0\\i \neq\{j,j+2\}}}^3 |i\>\<j|\otimes |j\>\<i|
   \nonumber\\
   &+& \left(a-\frac 12\right) \sum_{\substack{i=0\\i=j+2}}^3 |i\>\<j|\otimes |j\>\<i|.
\end{eqnarray}
 Clearly, for $\theta \in [0,\pi] - \{\pi/2\}$, one has $\frac 12 \leq a \leq 1,~0 \leq b,d \leq 1,~1 \leq c \leq \frac 32$. Hence, the first part of Eq.~\eqref{c2c2} is positive. Now we need to show that $B_2(\theta)\geq 0$. Note that the positivity of $B_2(\theta)$ is equivalent to the positivity of the following two $2\times 2$ submatrices
\begin{eqnarray}
  A_3=  \left(
\begin{array}{cc} \frac 32 - b & a - \frac 32 \\ a - \frac 32 & \frac 32 - d \end{array} \right)~~\mbox{and}~~
A_4=\left(
\begin{array}{cc} \frac 32 - c & a - \frac 12 \\ a - \frac 12 & \frac 32 - c \end{array} \right). \nonumber\\
\end{eqnarray}
Simple calculation shows that $\det[A_3]=\frac 12 (1-\sin\theta) \geq 0$ and  $\det[A_4]=0$ in $\theta \in  [0,\pi] - \{\pi/2\}$.
This proves that $B_2(\theta) \geq 0$ and hence $\tilde{W}^{\rm M}_{I}(\theta)$ is decomposable.


\begin{thebibliography}{10}

\bibitem{HHHH} R. Horodecki, P. Horodecki, M. Horodecki, and K. Horodecki, {\em Quantum entanglement}, Rev. Mod. Phys. {\bf 81}, 865 (2009).

\bibitem{EW1} B.M. Terhal, {\em Bell inequalities and the separability criterion}, Phys. Lett. A {\bf  271}, 319 (2000).

\bibitem{EW2} O. G\"uhne and G. T\'oth, {\em Entanglement detection}, Phys. Rep. {\bf 474}, 1 (2009).


\bibitem{TOPICAL} D. Chru\'sci\'nski and G. Sarbicki, {\em Entanglement witnesses: Construction, analysis and classification}, J. Phys. A: Math. Theor. {\bf 47}, 483001 (2014).

\bibitem{KYE}  S.-H. Kye, Facial structures for various notions of positivity and applications to the theory of entanglement, Rev. Math. Phys. {\bf 25}, 1330002 (2013).

\bibitem{ani18}
Anindita Bera, Shiladitya Mal, Aditi Sen(De), and Ujjwal Sen, {\it Witnessing bipartite entanglement sequentially by multiple observers}, Phys. Rev. A {\bf 98}, 062304 (2018).

\bibitem{Jam}
A. Jamio\l{}kowski, {\em Linear transformations which preserve trace and positive semidefiniteness of operators}, Rep. Math. Phys. {\bf 3}, 275 (1972).

\bibitem{MDChoi}
M.-D. Choi, {\em Completely positive linear maps on complex matrices}, Linear Algebra Appl. {\bf 10}, 285 (1975).

\bibitem{Paulsen} V. Paulsen, {\it Completely Bounded Maps and Operator
Algebras} (Cambridge University Press, Cambridge, 2003).

\bibitem{Stormer} E. St{\o}rmer, {\it Positive Linear Maps of Operator Algebras}, Springer Monographs in Mathematics (Springer, New York, 2013).

\bibitem{ani22}
Anindita Bera, Giovanni Scala, Gniewomir Sarbicki, Dariusz  Chru{\'s}ci{\'n}ski,
{\it Generalizing Choi map in $M_3$ beyond circulant scenario}, arXiv: 2212.03807.



\bibitem{Lew} M. Lewenstein, B. Kraus, J. I. Cirac, and P. Horodecki, {\em Optimization of entanglement witnesses}, Phys. Rev. A {\bf 62}, 052310 (2000).



\bibitem{SPA-01} P. Horodecki, {\em From limits of quantum operations to multicopy entanglement witnesses and state-spectrum estimation}, Phys. Rev. A {\bf 68}, 052101 (2003).

\bibitem{SPA-02} P. Horodecki and A. Ekert, {\em Method for direct detection of quantum entanglement}, Phys. Rev. Lett. {\bf 89}, 127902 (2002).



\bibitem{SPA-1} J. K. Korbicz, M. L. Almeida, J. Bae, and M. Lewenstein, A. Acin, {\em Structural approximations to positive maps and entanglement breaking channels}, Phys. Rev. A {\bf 78}, 062105 (2008). 

\bibitem{SPA-2} R.  Augusiak, J.  Bae, L. Czekaj, and M.  Lewenstein, {\em On structural physical approximations and entanglement breaking maps}, J. Phys. A: Math. Theor. {\bf 44}, 185308 (2011). 


\bibitem{SPA-3}  R. Augusiak, J. Bae, J. Tura, M. Lewenstein, {\em Checking the optimality of entanglement witnesses: an application to structural physical approximations},  J. Phys. A: Math. Theor. {\bf 47}, 065301 (2014). 

\bibitem{SPA-4} F. Shultz, {\em The structural physical approximation conjecture}, J. Math. Phys. {\bf 57}, 015218 (2016). 


\bibitem{EX1}  D. Chru\'sci\'nski, J. Pytel, and G. Sarbicki, {\em Constructing new optimal entanglement witnesses}, Phys. Rev. A {\bf 80}, 062314 (2009).

\bibitem{EX2} D . Chru\'sci\'nski and J. Pytel, {\em Optimal entanglement witnesses from generalized reduction and Robertson maps}, J. Phys. A: Math. Theor. {\bf 44},  165304 (2011).

\bibitem{S71} K.-C. Ha and S.-H. Kye, {\em One-parameter family of indecomposable optimal entanglement witnesses arising from generalized Choi maps}, Phys. Rev. A {\bf 84}, 024302 (2011).




\bibitem{hakyend} K.-C. Ha and S.-H. Kye,  {\em The structural physical approximations and optimal entanglement witnesses}, J. Math. Phys. {\bf 53}, 102204 (2012).

\bibitem{SPA-Stormer} E. St{\o}rmer, {\em Separable states and the structural physical approximation of a positive map}, J. Funct. Anal. {\bf 264}, 2197 (2013).

\bibitem{dar-gni} D. Chru\'sci\'nski and G. Sarbicki , {\em Disproving the conjecture on the structural physical approximation to optimal decomposable entanglement witnesses}, J. Phys. A: Math. Theor. {\bf 47}, 195301 (2014). 

\bibitem{MEW} J. Bae, D.  Chru\'sci\'nski, and B.C. Hiesmayr,  {\em Mirrored Entanglement Witnesses}, 	NPJ Quantum Inf. {\bf 6}, 15 (2020).



\bibitem{LAA} A. Bera, G. Sarbicki, and  D.  Chru\'sci\'nski,  {\em A class of optimal positive maps in $M_n$}, arXiv:2207.03821.

\bibitem{maciej00} M. Lewenstein, B. Kraus, J. I. Cirac, and P. Horodecki,  {\em Optimization of entanglement witnesses}, Phys. Rev. A {\bf 62}, 052310 (2000).

\bibitem{CES1} K. R. Parthasarathy,  {\em On the maximal dimension of a completely entangled subspace for finite level quantum systems}, Proc. Math. Sci.  {\bf 114}, 365 (2004).

\bibitem{CES2} T. Cubitt, A. Montanaro, and A. Winter,  {\em On the dimension of subspaces with bounded Schmidt rank}, J. Math. Phys. {\bf 49}, 022107 (2008).

\bibitem{jpa22}
 J. Bae, A. Bera, D. Chru{\'s}ci{\'n}ski, B. C. Hiesmayr, and D. McNulty,  {\it How many mutually unbiased bases are needed to detect bound entangled states}, J.  Phys. A: Math. Theor. {\bf 55}, 505303 (2022).

\bibitem{Lew-2} M. Lewenstein, B. Kraus, P. Horodecki, and J. I. Cirac,  {\em Characterization of separable states and entanglement witnesses}, Phys. Rev. A {\bf 63}, 044304 (2001).


 \bibitem{magic1} 
B. Baumgartner, B. C. Hiesmayr, and H. Narnhofer, {\em State space for two qutrits has a phase space structure in its core}, Phys. Rev. A {\bf 74}, 032327 (2006).  

\bibitem{magic2} 
B. Baumgartner, B. C. Hiesmayr, and H. Narnhofer, {\em Special simplex in the state space for entangled qudits}, J. Phys. A:
Math. Theor. {\bf 40}, 7919 (2007).

\bibitem{magic3}  
B. Baumgartner, B. C. Hiesmayr, and H. Narnhofer, {\em The geometry of bipartite qutrits including bound entanglement},
Phys. Lett. A {\bf 372}, 2190 (2008). 

\bibitem{MagicSimplex1}
B. C. Hiesmayr,  {\it Free versus Bound Entanglement: Machine learning tackling a NP-hard problem},
Sci. Rep. {\bf 11},  19739 (2021).

\bibitem{MagicSimplex2}
Ch. Popp and B. C. Hiesmayr,  {\em  Almost complete solution for the NP-hard separability problem of Bell diagonal qutrits},
Sci. Rep. {\bf 12},  12472 (2022).

\bibitem{hiesmayr3}
C. Popp and B. C. Hiesmayr,  {\it Bound Entanglement of Bell Diagonal Pairs of Qutrits and Ququarts: A Comparison}, arXiv:2209.15267.



\bibitem{UPB} C. H. Bennett, D. P. DiVincenzo, T. Mor, P. W. Shor, J. A. Smolin, and B. M. Terhal,  {\em Unextendible product bases and bound entanglement}, Phys. Rev. Lett. {\bf 82}, 5385 (1999).

\bibitem{UPB-1} B.M. Terhal,  {\em A Family of indecomposable positive linear maps based on entangled quantum states}, Linear Algebra  Appl. {\bf 323}, 61 (2001).


\bibitem{Korea-1992} S. J. Cho, S.-H. Kye, and S. G. Lee,  {\em Generalized Choi maps in 3-dimensional matrix algebra}, Linear Algebra Appl. {\bf 171}, 213 (1992).



\bibitem{Choi-2} M. D. Choi,  {\em Positive semidefinite biquadratic forms}, Linear Algebra Appl. {\bf 12}, 95 (1975).


\bibitem{Choi-4} M. D. Choi,  {\em Positive linear maps}, Proc. Symp. Pure Math. {\bf 38}, 583 (1982).

\bibitem{Choi-5} M. D. Choi and T. Y. Lam,  {\em Extremal positive semidefinite forms}, Math. Ann. {\bf 231}, 1 (1977).

\bibitem{Ha-extr} K.-C. Ha,  {\em Notes on extremality of the Choi map}, Linear Algebra Appl. {\bf 439}, 3156 (2013).

 \bibitem{S72}  D. Chru\'sci\'nski and G. Sarbicki, {\em Optimal entanglement witnesses for two qutrits}, Open Sys. Inf. Dyn. {\bf 20}, 1350006 (2013).


\bibitem{Kossak} A. Kossakowski,  {\em A Class of Linear Positive Maps in Matrix Algebras}, Open Sys. Inf. Dyn. {\bf 10}, 213 (2003).

\bibitem{FilipI} D. Chru\'sci\'nski and F. A. Wudarski,  {\em Geometry of entanglement witnesses for two qutrits}, Open Syst. Inf. Dyn. {\bf  18}, 387 (2011).





\bibitem{PRA-2022} A. Bera, F. A. Wudarski, G. Sarbicki and D. Chru\'sci\'nski, {\em  Class of Bell-diagonal entanglement witnesses in $\mathbb{C}^4 \otimes \mathbb{C}^4$: Optimization and the spanning property}, Phys. Rev. A {\bf 105}, 052401  (2022).


\bibitem{Breuer1}
H. -P. Breuer,  {\em Optimal Entanglement Criterion for Mixed Quantum States}, Phys. Rev. Lett. {\bf 97}, 0805001 (2006).

\bibitem{Hall1}
W. Hall,  {\em A new criterion for indecomposability of positive maps}, J. Phys. A: Math. Gen. {\bf 39}, 14119 (2006).



\bibitem{TT} K. Tanahashi and J. Tomiyama,  {\em Indecomposable positive maps in matrix algebras}, Can. Math. Bull. {\bf 31}, 308 (1988).

\bibitem{Osaka1} H. Osaka,  {\em Indecomposable Positive Maps in Low Dimensional Matrix Algebra}, Linear Algebra Appl. {\bf 153}, 73 (1991).

\bibitem{Osaka2} H. Osaka,   {\em A Series of Absolutely
Indecomposable Positive Maps in Matrix Algebras}, Linear Algebra Appl. {\bf 186}, 45 (1993).

\bibitem{Ando} T. Ando,  {\em Positivity of certain maps}, Seminar Notes, 1985 (cited in [31]).

\bibitem{RIMS} K.-C. Ha,  {\em Atomic positive linear maps in matrix algebras}, Publ. RIMS, Kyoto Univ. {\bf 34}, 591 (1998).

\bibitem{yamagami}  S. Yamagami,  {\em Cyclic inequalities},  Proc. Am. Math. Soc. {\bf 118}, 521 (1993).







\bibitem{EnglertBound} W. Li, R. Han, J. Shang, H.K. Ng, and B.-G. Englert, 
 {\it Sequentially constrained Monte Carlo sampler for quantum states},
 arXiv:2109.14215.

\bibitem{CPHiesmayr}
C. Spengler, M. Huber, and B. C. Hiesmayr,  {\it Composite parameterization and Haar measure for all unitary and special unitary groups}, J. Math. Phys. {\bf 53}, 013501 (2012).

\bibitem{HiesmayrLorentzboost} 
P. Caban and B. C. Hiesmayr, {\it Is bound entanglement Lorentz invariant?}, 
 arXiv:2212.01286.


\end{thebibliography}
\end{document}